\documentclass[titlepage,12pt]{article}
\usepackage[margin=1in]{geometry}
\usepackage{latexsym}
\usepackage{amsmath,amsthm,natbib}
\usepackage{latexsym, epsfig, verbatim,  amsmath, amssymb, lscape,multirow}
\usepackage{amsfonts}
\usepackage{hyperref,setspace,soul}
\usepackage{color}
\hypersetup{
  colorlinks = true,
  urlcolor = black
}
\usepackage{graphicx}
\newtheorem{remark}{Remark}

\newcommand{\pr}{{P}}
\newcommand{\cond}{{|}}

\newcommand{\V}{\mathcal{V}}
\newcommand{\D}{{d}}
\newcommand{\T}{{\intercal}}
\newcommand{\pn}{{\mathbb{P}_n}}

\newcommand{\argmin}{\mbox{argmin}}
\newcommand{\sign}{\mbox{sgn}}

\begin{document}

\title{\bf \Large {Constructing Stabilized Dynamic Treatment Regimes for Censored Data}}

 \author{Ying-Qi
   Zhao \thanks{Associate Member, Public Health Sciences Division, Fred Hutchinson Cancer Research Center, Seattle, WA, 98109, Email: yqzhao@fhcrc.org.}
   and Ruoqing Zhu \thanks{Assistant Professor, Department of Statistics, University of Illinois Urbana-Champaign, Champaign, Illinois, 61820}
   and Guanhua Chen \thanks{Assistant Professor, Department of Biostatistics and Medical Informatics, University of Wisconsin-Madison, Madison, Wisconsin, 53792}
   and Yingye Zheng \thanks{Member, Public Health Sciences Division, Fred Hutchinson Cancer Research Center, Seattle, Washington, 98109}
   }

\date{ \today}
\maketitle

\begin{abstract}
Stabilized dynamic treatment regimes are sequential decision rules for individual patients that not only adaptive throughout the disease progression but also remain consistent over time in format. The estimation of stabilized dynamic treatment regimes becomes more complicated when the clinical outcome of interest is a survival time subject to censoring. To address this challenge, we propose two novel methods, censored shared-Q-learning and censored shared-O-learning. Both methods incorporate clinical preferences into a qualitative rule, where the parameters indexing the decision rules are shared across different stages and estimated simultaneously. We use extensive simulation studies to demonstrate the superior performance of the proposed methods. The methods are further applied to the Framingham Study to derive treatment rules for cardiovascular disease. \\

\end{abstract}

\section{Introduction}
Dynamic treatment regimes (DTRs), also called adaptive treatment strategies \citep{Murphy03optimaldynamic, Murphy:ExperimentalDesign2005}, are sequential decision rules adapting over time to the time-varying characteristics of patients. A DTR takes patient health histories as inputs and recommends the next treatment strategy at each decision point. For example, treatment for lung cancer usually involves regimens with multiple lines \citep{Socinski:NSCLC2007}. Clinicians may update treatment for major depressive disorder according to factors emerging over time, such as side-effect severity, treatment adherence, and so on \citep{Murphy:MethodologiclChallengesPsychiatricDisorders07}. In these examples, the decision rules are different across different stages, yet in practice, it is not unusual  to have a common decision rule shared across different stages. For example, diabetes patients are recommended  medication or lifestyle intervention when Hemoglobin A1c (HbA1c) rises above a threshold,  which is universal across the entire medication process. Lung transplantation may be initiated for a cystic fibrosis patient if FEV$_1\%$ (Forced Expiratory Volume in 1 second) value falls below 30\%, which, again, remains the same throughout the course of disease progression.  Such shared decision rules are easier to implement in practice throughout multiple decision points, in particular, when multivariate time-varying covariates are potentially involved. We term them as stabilized dynamic treatment regimes (SDTRs), which are also referred to as DTRs with shared parameters \citep{chakraborty2016estimating}. We will use both terms interchangeably throughout the article. 

Estimating an optimal DTR without shared parameters has been widely studied in the past few years.
A well-established approach is Q-learning \citep{watkins1989learning, nahum2012q, zhao2009reinforcement, laber2014interactive, goldberg2012q}, which recursively estimates the conditional expectation of the outcomes given the current patient history, assuming that optimal decisions are made in the future. The foregoing conditional expectations are known as Q-functions. Semiparametric methods have also been proposed, such as iterative minimization of regrets \citep{Murphy03optimaldynamic} and G-estimation \citep{robins2004optimal}. These methods are methodologically more complex, but could potentially provide efficiency gain. Recently, direct methods have become popular in the  literature. They circumvent the modeling of conditional means of the outcome given the treatment and covariates, and they directly estimate the decision rule that maximizes the expected outcomes. Examples include backward outcome weighted learning, simultaneous outcome weighted learning \citep{zhao2015new}, and others \citep{robins2008estimation, orellana2010dynamic, zhang2013robust}.  
We refer readers to \citet{chakraborty2013statistical} and \citet{kosorok2015adaptive} for detailed reviews of the current literature.

SDTRs are analogous to stationary Markov decision processes with function approximations \citep{sutton1998reinforcement}. For solving Markovian decision problems, \citet{antos2008learning} proposed to minimize the squared Bellman error, where Bellman error  quantifies  the difference between the estimated reward at any time point, and the actual reward received. \citet{chakraborty2016estimating} proposed shared Q-learning to estimate the optimal shared-parameter DTR when the decision rule at each stage is the same function of one or more time-varying covariates. In particular, they formulated the decision rules as linear functions of covariates, and the coefficients are assumed to be the same across stages. An alternative way of identifying an SDTR is the simultaneous G-estimation \citep{robins2004optimal}, which can handle problems with shared parameters in principle, but its empirical performance is largely unknown \citep{moodie2010estimating}. The simultaneous outcome weighted learning approach proposed in \citet{zhao2015new} (with some modification) could be used when the goal is to derive SDTRs. They converted the construction of DTRs into a simultaneous nonparametric  classification problem, where a multi-dimensional hinge loss  of the expected long-term outcome was employed. However, they did not explore modifying the method for SDTRs.

Moreover, none of the aforementioned methods can handle censored survival outcome when constructing an SDTR. In general, methods for accommodating time-to-event outcomes are mostly limited to the regular DTR settings. This is mostly due to two main challenges. First, the number of stages for each individual in the study is not fixed. This is because the event time can vary by individual, and the treatment is usually stopped once the failure event happens. Second, the treatment/outcome status of a subject may be unknown when censoring occurs. To this end, \citet{goldberg2012q} developed  a Q-learning algorithm  to adjust   for censored data and allow a flexible number of stages. However, it cannot be directly applied to solve for SDTRs.

In this paper, we propose two methods  for solving SDTRs, termed as censored shared-Q-learning and censored shared-O-learning (abbreviated from ``outcome weighted learning''). The censored shared-Q-learning method generalizes the shared-Q-learning method \citep{chakraborty2016estimating} by applying the inverse probability of censoring weights to account for the uncertainty in the outcomes of censored subjects, where the censoring weights need to be carefully constructed for each stage due to the multi-stage nature of the problem. Similar to shared Q-learning, censored shared-Q-learning is an iterative procedure, which identifies the optimal decision rule for each stage in a sequential/iterative manner.  The censored shared-O-learning method  uses a non-iterative approach by directly maximizing a concave relaxation of the inverse-probability-of-censoring weighted estimator of the expected survival benefit. To the best of our knowledge, this is the first article that provides a thorough solution for deriving DTRs  with shared parameters in the censored data setup.

The remainder of the paper is organized as follows. We introduce the general framework of SDTR in Section \ref{scn_framework}.  In Section \ref{scn_method}, we introduce censored shared-Q-learning and censored shared-O-learning along with the computation algorithms.  We conduct numerical studies that compare the proposed methods with  Q-learning and shared Q-learning in Section \ref{scn_sim}. Section \ref{scn_data} focuses on the application of the proposed methods to the Framingham Heart data. Finally, we provide a discussion of open questions in Section \ref{scn_disc}.

\section{Statistical framework}\label{scn_framework}
In this section, we present definitions and notations used in the paper, where we follow \cite{goldberg2012q} whenever possible. Throughout, we use uppercase letters to denote random variables and lowercase letters to denote realizations of the random variables.

Let $T$ be the maximal number of stages in a multistage study.  Here, stages are referred to as clinical decision time points. A full trajectory of an observation  sequence is  $\{X_{1}, A_1, Y_1, \ldots, X_T, A_T, Y_T\}$. For any decision point $j = 1, \ldots, T$, the information of a patient is represented by $(X_j, A_j, Y_j)$, where $X_1$ denotes the initial information, $X_j$ denotes the intermediate information collected between stages $j-1$ and $j$ when $j > 0$, and $A_j$ is the treatment assigned at the $j^{th}$ stage subsequent to observing $X_j$. We assume that there are two possible treatments at each stage, i.e. $A_j \in \mathcal{A}_j = \{-1,1\}$. $Y_j$ is a non-negative random variable that equals to either the length of the interval between decision time point $j$ and $j+1$, or the length between decision time point $j$ and failure time if the failure event occurs in that stage. $Y_j$ can be viewed as the reward in the $j^{th}$ stage, and $\sum_{k= 1}^j Y_k$ is the cumulative reward  up to and including stage $j$. In our context, the sum $\sum_{k= 1}^j Y_k$ is the total survival time up to and including stage $j$.   As the total survival times are broken down based on the number of stages, this setup could introduce complexity to the trajectory structure. In particular, if a failure event occurs before the final decision point $T$, the trajectory will not be of full length \citep{goldberg2012q}. Subsequently, the number of stages for different observations could be different. We denote the (random) total number of stages for an individual by $\bar T$. Thus, $\sum_{j= 1}^{\bar T} Y_j$ is the overall survival time. 

In this study, the observations are subject to censoring. Let $C$  denote the censoring time, taking values in the segment $[0, \tau]$. We assume that the censoring and failure times are independent given the covariates of all previous stages. Let $\Delta_j$ be the censoring indicator at stage $j$, where $\Delta_j = I(\sum_{k=1}^j Y_k \leq C)$. If no censoring event happens before the $(j+1)^{th}$ decision time point, then $\Delta_j = 1$ and the outcome $Y_j$ is observed.  Hence, if a censoring event occurs during stage $\bar T$, then $C \leq \sum_{j=1}^{\bar T} Y_j$, $\Delta_{\bar T -1} = 1$ and $\Delta_{\bar T} = 0$. In this case, we cannot observe the failure time that is censored by $C$. On the other hand, $\Delta_{\bar T} = 1$ if the failure time $ \sum_{j=1}^{\bar T} Y_{j}$ is observed. While $\Delta_j = 1$ indicates that censoring has not occurred at stage $j$, it does not necessarily mean $\sum_{k=1}^j Y_k$ is the time of event, which is different from the notation of a typical survival analysis. Only $\Delta_{\bar T}$ is the true failure time indicator. 

We use an overbar to denote the collection of history information, e.g. the sequence of treatments $(A_1, A_2,..., A_j)$ is represented by $\bar A_j$, and $\bar X_j$ is the sequence of covariates up to $j$. Let $H_j=(\bar X_j, \bar Y_{j-1}, \bar A_{j-1}) \in \mathcal{H}_j$ denote the accrued information at each stage, with the convention that $A_0 = \emptyset$, $Y_0 = 0$ and $H_1=X_1 \in \mathcal{H}_1$. A DTR is a sequence of deterministic decision rules, $d=(d_1,\ldots,d_T)$, where $d_j: \mathcal{H}_j \mapsto \mathcal{A}_j$ is a function mapping from the space of accrued information to the space of available treatments. Under $d$, a patient presenting with $H_j = h_j$ at time $j$ is recommended to treatment $d_j(h_j)$. We use $\mathcal{D}_T$ to denote the class of all possible treatment regimes. Our goal is to identify the optimal dynamic treatment regime $d^* \in \mathcal{D}_T $, that maximizes the expected outcome if deployed to the whole population in the future.  
Given that no information on survival is available beyond $\tau$ in the observed data, the outcome of interest is truncated-by-$\tau$ expected survival. Let $P^d$ denote the distribution of a trajectory given that $A_j = d_j(H_j), j = 1, \ldots, \bar T$. The optimal DTR is the sequence of rules that maximizes
\begin{equation}
E^d\Bigg\{\min\bigg(\sum_{j=1}^{\bar T} Y_j, \tau\bigg)\Bigg\},
\label{sharedQobj1}
\end{equation}
where $E^d$ is the expectation with respect to $P^d$.  We refer to \ref{sharedQobj1}) as the value associated with a regime $d$, denoted by $\mathcal{V}(d)$.

Following \citet{goldberg2012q}, we modify the trajectories when they are not of full length or the overall survival time is greater than $\tau$. The idea is to extend the information to full length by introducing noninformative values at stages after the failure. If a failure time  occurs at stage $j < T$, we let $X_k' = X_k$ for $1 \leq k \leq j+ 1$ and let the noninformative $X_k' = \emptyset$, for $j +1 < k \leq T$. Similarly, we let $A_k' = A_k$ and $Y_k' = Y_k$, for $1 \leq k \leq j$. And for $j + 1 \leq k \leq T$, we set $Y_k' = 0$ and draw $A_k'$ uniformly from $\mathcal{A}$ as the noninformative values. If for some $j$, $\sum_{k}^j Y_j \geq \tau$, then we modify $Y_j'$ to be $Y_j' = \max(\tau - \sum_{k=1}^{j-1} Y_j, 0)$, and modify the trajectory to be noninformative at all stages after $j$. For any DTR $d \in \mathcal{D}_T$, we define a corresponding $d'$ for the modified trajectories, where the same action is chosen for any triplet $(x_j', a_{j-1}', y_{j-1}')$ if $x_j' \neq \emptyset$, and a fixed action is chosen if $x_j' = \emptyset$. It has been shown in \citet{goldberg2012q} that
\begin{equation}
E^{d'}\bigg(\sum_{j=1}^{T} Y_j'  \Big| X_1 = x_1 \bigg)  = E^d\Bigg\{\min\bigg(\sum_{j=1}^{\bar T} Y_j, \tau\bigg) \Big| X_1 = x_1\Bigg\}.
\label{value}
\end{equation}
Subsequently, $\mathcal{V}(d)$ remains the same after this modification. In the following, we omit the ``prime'' in the modified trajectories without the risk of ambiguity. Another complication is due to censoring, where the trajectories themselves may be censored and cannot be fully observed. This needs to be carefully handled when estimating SDTRs from the data.

\section{Estimating DTRs with shared parameters for censored data } \label{scn_method}
We often encounter time-varying covariates in longitudinal studies.  To facilitate clinical implementation, it could be beneficial to have a shared decision rule in its functional form across multiple stages, while allowing the covariate values to change over time. In the following, we propose two methods to estimate the DTRs with shared parameters using repeatedly measured covariate information when the outcome is subject to censoring. The proposed algorithms are based on Q-learning and O-learning in the multistage setup, respectively.

\subsection{Censored shared-Q-learning} \label{scn_sharedQ}

Due to delayed effects, we need to consider the entire treatment sequence in order to optimize the long-term outcome. Results from the dynamic programming literature show that $d_{T}^{*}(h_T) = \arg\max_{a_T}Q_T(h_T, a_T)$, where $Q_T(h_T, a_T) = E(Y_T \cond H_T=h_T, A_T=a_T)$ and recursively
$d_{j}^*(h_j) = \arg\max_{a_j}Q_{j}(h_j, a_j)$ where $Q_j(h_j, a_j) = E(Y_j + \max_{a_{j+1}}Q_{j+1}(H_{j+1}, a_{j+1})\cond H_j=h_j, A_j=a_j)$ for $j=T-1,\ldots,
1$ when the underlying generative distribution is known~\citep{Bellman:DynamicProgramming09,Sutton:Reinforcementlearning98}. Q-learning is an approximate dynamic programming algorithm that uses regression models to estimate the Q-functions $Q_{j}(h_j, a_j), j=1,\ldots, T$ and then estimate $d_{j}^*(h_j)$ recursively. Note that if a failure occurred prior to the $j^{th}$ time point, i.e., $X_j = \emptyset$, we set $Q_j$ as zero. A commonly used strategy is to estimate Q-functions via linear regression using working models $Q_j(H_j, A_j; \beta_j, \psi_j) = \beta_j^\T H_{j0} + (\psi_j^\T H_{j1})A_j$, where $H_{j0}$ and $H_{j1}$ (including intercepts) are possibly different features of $H_j$.

We focus on the setting where the decision rule parameters are shared across stages with $\psi_1 = \ldots = \psi_T = \psi$, but $\beta_j$s are left unshared. Let $\theta_j = (\beta_j^\T, \psi)^\T$. Assuming the shared model is correctly specified, the pseudo-outcome at the $j^{th}$ stage is $\tilde Y_j(\theta_{j+1}) = Y_j +  \beta_{j+1}^\T H_{j+1, 0} + |\psi^\T H_{j+1, 1}|$, and $Q_j(H_j, A_j) = E\{  \tilde Y_j(\theta_{j+1}) | H_j, A_j \}, j = 1, \ldots, T-1$. Let $\tilde Y_T(\theta_{T+1}) = Y_T$. For any stage $j$, we can write $ Q_j( H_j, A_j) = \beta_j^\T H_{j0} + (\psi^\T H_{j1})A_j$. Hence,  $(\widehat \beta_1, \ldots, \widehat\beta_T, \widehat\psi)$ can be estimated \citep{chakraborty2016estimating} via
\begin{equation}
\underset{\beta_1,\ldots, \beta_T, \psi}{\argmin} \,\, \sum_{j=1}^\T \pn\left [\tilde Y_j(\theta_{j+1})- \{\beta_j^\T H_{j0} + (\psi^\T H_{j1})A_j \} \right]^2,
\label{sharedQobj}
\end{equation}
where $\pn$ denotes  the empirical averages of the data.  However, $Y_j$ in the $\tilde Y_j(\theta_{j+1})$ may be censored and unknown in the censoring data setup, and $\tilde Y_j(\theta_{j+1})$s are defined by unknown parameters.

Let $S_C(\sum_{k=1}^j Y_k \cond H_{j }, A_{j}) = \pr(C>\sum_{k=1}^j Y_k \cond H_j, A_{j})$ be the conditional survival function for the censoring time given history information up to stage $j$ and treatment received at that stage. In addition, we assume the conditional independent of censoring (i.e. $\sum_{k=1}^j Y_k \perp C \cond H_j, A_j$). Then
\begin{eqnarray*}
  E\bigg\{\frac{ \Delta_j}{ S_C\big\{\min(\sum_{k=1}^j Y_k, C) \cond H_j, A_j\big\}} \Big| H_j, A_j \bigg\} & = & 1.
\end{eqnarray*}
Let $U_j = \min(\sum_{k=1}^j Y_k, C) - \sum_{k=1}^{j-1} Y_k $ if the censoring occurs in the $j^{th}$ stage, and $U_j = Y_j$ otherwise. We further define $\tilde U_j(\theta_{j+1}) = U_j +  \beta_{j+1}^\T H_{j+1, 0} + |\psi^\T H_{j+1, 1}|$ for $j \leq T-1$ and  $\tilde U_T(\theta_{T+1}) = U_T$. Consequently,
\begin{eqnarray*}
  && E\left [\tilde Y_j(\theta_{j+1})  - \{\beta_j^\T H_{j0} + (\psi^\T H_{j1})A_j \} \right]^2 \\
  & = & E\left(\left [\tilde U_j(\theta_{j+1})  - \{\beta_j^\T H_{j0} + (\psi^\T H_{j1})A_j  \}\right]^2 \frac{ \Delta_j}{ S_C\{\min(\sum_{k=1}^j Y_k, C) \cond H_j, A_j\}}\right).
\end{eqnarray*}
The quantity reflected in the above equation will only involve observed data, thus instead of \eqref{sharedQobj}, $(\widehat\beta_1^\T, \ldots, \widehat\beta_T^\T, \widehat\psi)^\T$ can be estimated via a weighted least square procedure, with
\begin{eqnarray}
\underset{\beta_1,\ldots, \beta_T, \psi}{\argmin} \,\, \sum_{j=1}^\T \pn \left(\left [\tilde U_j(\theta_{j+1})- \{\beta_j^\T H_{j0} + (\psi^\T H_{j1})A_j \} \right]^2 \frac{ \Delta_j}{\widehat S_C\{\min(\sum_{k=1}^j Y_k, C) \cond H_j, A_j\}}\right),
\label{sharedQ}
\end{eqnarray}
where $\widehat S_C$ is an estimator for $S_C$.


We employ an iterative procedure for estimating $\theta_j$s, which was also applied in \citet{chakraborty2016estimating} for shared Q-learning without censoring.  The censored shared-Q-learning algorithm is presented  below.
\begin{enumerate}
\item Estimate the conditional survival function for the censoring time $\widehat S_C\{\min(\sum_{k=1}^j Y_k, C) \cond H_j, A_j\}$ at stage $j$, and construct the diagonal matrix
$$
V = \left(
\begin{array}{cccc}
 V_T & 0 & \ldots & 0 \\
 0 & V_{T-1} & \ldots & 0 \\
 \vdots & \vdots & & \vdots\\
 0 & 0 & \ldots & V_1 \\
\end{array}
\right)
$$
where $V_j =  { \Delta_j}/{\widehat S_C\{\min(\sum_{k=1}^jY_k, C) \cond H_j, A_j\}}, j = 1, \ldots, T$.
\item Set the initial value of $\theta$, denoted by $\widehat\theta^{(0)} = \left( \widehat\beta_T^{{(0)}^\T}, \ldots,  \widehat\beta_1^{{(0)}^\T},  \widehat\psi^{{(0)}^\T} \right)^\T$.
\item At the $(l+1)^{th}$ iteration, $l = 0, 1, 2, \ldots$:
\begin{enumerate}
\item Constructing the vector $\tilde U(\widehat \theta^{(l)}) = \left(\tilde U_T({\widehat\theta_{T+1}^{(l)})}^\T, \ldots, \tilde U_1({\widehat\theta_{2}^{(l)})}^\T\right)^\T$.

\item Solving for $\widehat\theta^{(l + 1)} = (Z^{\T} V Z)^{-1} Z^\T V \tilde U(\widehat \theta^{(l)})$, where
$$
Z = \left(
\begin{array}{ccccc}
 H_{T0}^\T & 0 & \ldots & 0  & H_{T1}^\T A_T\\
 0 & H_{T-1,0}^\T  & \ldots & 0 & H_{T-1, 1}^\T A_{T-1} \\
 \vdots & \vdots & & \vdots & \vdots\\
 0 & 0 & \ldots & H_{10}^\T & H_{11}^\T A_1\\
\end{array}
\right).
$$
\end{enumerate}
\item Repeat steps (3a)-(3b) until $\|\widehat\theta^{(l + 1)} - \widehat\theta^{(l )} \| \leq \epsilon$ for a prespecified value $\epsilon$ or until the maximum number of iterations is reached.
\end{enumerate}

A simple  choice  for  the initial values of $\theta$ is to sets all parameters to zero, $\widehat\theta^{(0)} = (0^\T, \ldots, 0^\T, 0^\T)^\T$. In this paper, we let the initial values depend  on the estimates from censored Q-learning \citep{goldberg2012q}, where the parameters are not shared. Denote the estimates as $\widehat\beta_1^{(0)}, \ldots, \widehat\beta_T^{(0)}, \widehat\psi_1^{(0)}, \ldots, \widehat\psi_T^{(0)}$. We combine the distinct estimates of $ \widehat\psi_j^{(0)}$ into a single estimate via an average. The initial values of $\theta$ can be set as  $\widehat\theta^{(A)}=
 \left( \widehat\beta_T^{{(0)}^\T}, \ldots,  \widehat\beta_1^{{(0)}^\T},  \widehat\psi^{{(A)}^\T} \right)^\T$, where $\widehat\psi^{{(A)} } = \sum_{j=1}^T \widehat\psi_j^{(0)}/ T$.

\subsection{Censored shared-O-learning}
The censored shared-Q-learning algorithm requires correct specifications of Q-functions at each stage. This could be unrealistic in the multistage setup since  the underlying data generating mechanism is usually complicated. In this section, we propose censored shared-O-learning, which constructs the shared decision rules by directly targeting the overall benefit of the decision rule.

Define $\pi_j(a_j; h_j) = \pr(A_j=a_j | H_j = h_j)$ for $j = 1,\ldots, T$.   We assume the following conditions, including i) $A_j$ is independent of all potential values of the outcome and future variables conditional on $H_j, j = 1, \ldots, T$; and (ii) $\pi_j(1; H_j)$ is strictly between 0 and 1. Assumption (i) is true in a sequential multiple assignment randomized trial \citep{murphy2005experimental} but unverifiable in an observational study. As shown in \citet{zhao2015new},
\begin{equation}
\V(d) = E^{d}\left(\sum_{j=1}^{T} Y_j  \right)   =  E\left[\frac{ (\sum_{j=1}^T Y_j)\prod_{j=1}^T I \{A_j=d_j(H_j)\}}
    {\prod_{j=1}^T \pi(A_j; H_j)}\right],
    \label{vd}
\end{equation}
where  $I(\cdot)$ is the indicator function. The right-hand side indicates $\V(d)$ equals the weighted average of outcomes among those who received $T$  treatments coinciding with that dictated by $d$, with weights $\{\prod_{j=1}^T \pi_j(A_j; H_j)\}^{-1}$. $\pi_j$s are usually known in a sequential multiple assignment randomized trial. If they are unknown, we can estimate  $\pi_j$s using methods such as a logistic regression. Given the estimates  $\widehat\pi_j(a_j; h_j)$, a plug-in estimator for $\V(d)$ based on \eqref{vd} is
$$
\widehat{\mathcal{V}}(d) =   \pn \left[\frac{ (\sum_{j=1}^T Y_j)\prod_{j=1}^T I \{A_j=d_j(H_j)\}}
    {\prod_{j=1}^T \widehat\pi(A_j; H_j)}\right].
$$
However, $\sum_{j=1}^T Y_j$ may not be fully observed due to censoring. Following the notion in Section \ref{scn_sharedQ}, let $U_T = \min (\sum_{j=1}^T Y_j, C)$ and $S_C(\sum_{j=1}^T Y_j \cond H_{T }, A_{T}) = \pr(C>\sum_{j=1}^T Y_j \cond H_T, A_{T})$ be the conditional survival function for the censoring time given history information up to stage $T$.  Denote the estimator of $S_C$ as $\widehat S_C$. We can estimate  $\V(d)$ in the scenario with time-to-event outcomes using
\begin{eqnarray}
\widehat \V(d) &= &\pn\left(\frac{U_T \prod_{j=1}^T I \{A_j=d_j(H_j)\}}{ \prod_{j = 1}^{T} \widehat\pi(A_j; H_j)  }\cdot \frac{\Delta_T}{\widehat S_C(U_T \cond H_{T}, A_{T})}  \right). \label{VestCensor}
\end{eqnarray}

The decision rules are formulated as fixed linear functions of the present variables at each stage. Mathematically, they are presented as  $d_j(H_j) = \sign(\psi^\T H_{j1})$, where $H_{j1}$ is a subset of $H_{j}$ important for determining the SDTRs similar to the one defined in Section $3.1$. In addition, we define $\sign(0) = 1$. Consequently, we maximize over $\psi$
\begin{eqnarray*}
\widehat \V(\psi) &=& \pn\left[ I \{\min_{j = 1, \ldots, T} A_j\psi^\T H_{j1} \geq 0  \} \frac{U_T \Delta_T}{ \prod_{j = 1}^{T} \widehat\pi(A_j; H_j)  \widehat S_C(U_T \cond H_{T}, A_{T})}  \right].
\end{eqnarray*}
where we use $\widehat \V(\psi)$ to denote $\widehat \V\{\sign(\psi^\T H_{j1})\}$, and substitute $I\{A_j = \D_j(H_j)\}$ by $I\{A_j\psi^\T H_{j1} \geq 0\}$.

It could be challenging to optimize  $\widehat \V(\psi)$ directly due to the discontinuity of the indicator functions.  A computationally efficient approach is to replace the indicator by a concave surrogate. This leads to an optimization problem of
\begin{equation}
\max_{(f_1, \ldots, f_T)} \pn\left(  \phi\left [\min_{j=1,\ldots, T}\left \{A_j \psi^\T H_{j1}\right\}\right] \frac{U_T \Delta_T}{ \prod_{j = 1}^{T} \widehat\pi(A_j; H_j)  \widehat S_C(U_T \cond H_{T}, A_{T})}   \right),
 \label{emp_obj}
\end{equation}
where $\phi$ is a concave function. In this paper, we will use $\phi(t)= -\log(1+e^{-t})$, which is an analog of the logistic loss in the machine learning literature. However, other choices of $\phi(t)$ are available; for example, analogs of exponential loss, hinge loss and others can also be applied \citep{bartlett2006convexity}.  The above objective function is not differentiable in $\psi$. To account for the non-differentiability of the minimum function in  \eqref{emp_obj}, we instead consider a soft-minimum function of $u$ and $v$ to replace $\min(u,v)$, which equals to
$
 -\log\{\exp(-uK) + \exp(-vK)\}/K,
$
with $K$ being a positive constant. Hence, the term $\min_{j=1,\ldots, T}\left \{A_j \psi^\T H_{j1}\right\}$, in \eqref{emp_obj} can be replaced by its soft-minimum counterpart. We maximize
\begin{eqnarray*}
 \Phi(\psi) &= &\pn \left( \frac{U_T \Delta_T}{ \prod_{j = 1}^{T} \widehat\pi(A_j; H_j)  \widehat S_C(U_T \cond H_{T}, A_{T})}  \log\Big[1 + K^{-1}\cdot  \sum_{j=1}^T \exp\left\{-K (A_j\psi^\T H_{j1})\right\}   \Big]\right).
\end{eqnarray*} 
The derivative with respect to shared parameters of the decision rules $\psi$ can be written as
\begin{eqnarray*}
	\frac{\partial \Phi(\psi)}{\partial \psi}  
	&= &\pn \left( \frac{U_T \Delta_T}{ \prod_{j = 1}^{T} \widehat\pi(A_j; H_j)  \widehat S_C(U_T \cond H_{T}, A_{T})}  
	\frac{- \sum_{j=1}^T \exp\left\{-K (A_j\psi^\T H_{j1})\right\} 	\cdot (A_{j} H_{j1}) }{1 + K^{-1}\cdot  \sum_{j=1}^T \exp\left\{-K (A_j\psi^\T H_{j1})\right\} } \right).
\end{eqnarray*}
We can employ the orthant-wise limited-memory quasi-Newton algorithm proposed by \cite{andrew2007scalable}. We also note that it is possible to write out the Hessian matrix for our proposed objective function. However, we find it does not benefit the numerical performance since the calculation of the second derivative is rather complicated and less efficient than a numerical approximation using the Sherman--Morrison updating formula, as implemented in the Broyden--Fletcher--Goldfarb--Shanno (BFGS) algorithm. 

Censored shared-O-learning maximizes the estimated mean outcome of a DTR over the pre-specified class of DTRs with shared parameters. Hence, compared with censored shared-Q-learning, it circumvents the need for estimating Q-functions at each stage. Furthermore, it does not require modeling the censoring distribution at each stage but only needs to model the censoring distribution at the final stage. However, censored shared-O-learning involves an unknown parameter $K$, which controls the approximation of the soft minimum function to the minimum function. In practice, we can use  cross-validation to select the best $K$ by grid search over a prespecified set of candidate values.

\begin{remark}
In practice, interpretable and simple rules are preferable. Sparse penalty such as LASSO can be applied in both censored shared-Q- and O-learning, where the coefficients for unimportant variables will shrink to zero. For  censored shared-Q-learning, we can solve for $\widehat\theta^{(l + 1)}$ in step 3(b) using penalized weighted least squares. For  censored shared-O-learning, we can maximize the penalized objective
$$ \Phi(\psi) - \tau_n  \|\psi\|_1,$$ where $\|\psi\|_1$ is the $L_1$ norm of $\psi$ and $\tau_n$ is a tuning parameter controlling the amount of penalization. The  orthant-wise limited-memory quasi-Newton algorithm, which is a limited-memory BFGS algorithm that incorporates the $\ell_1$ regulation, can still be applied in this case.
\end{remark}

\section{Simulation Studies} \label{scn_sim}

One of the motivations for the current work derives from  the long-term care of patients with diabetes. Patients are routinely examined for glycosylated hemoglobin (A1c) level every three months, and treatments are recommended for tightly controlling A1c to prevent adverse events such as hospitalization due to the disease.  Our simulation mimics such a setting, using  the generative model similar to that of  \citet{timbie2010diminishing} and \citet{ertefaie2018constructing}. We treat each check-up time as a decision point for determining  treatment in the next three months. Our study consists of 10 decision points.  The  treatments include metformin, sulfonylurea, glitazone, and insulin. Patients start with metformin and augment with treatments sulfonylurea, glitazone, and insulin during the study period. At each decision point, patients can either continue the current treatment or augment the treatment. A binary discontinuation indicator  is generated to represent patients' intolerance to treatment due to  side effects, and patients who discontinue a treatment will take the next available treatment. $N_j$ is the number of augmented treatments by the end of interval $j$ where $N_j \in \{0, 1, 2, 3, 4\}$, and the number of augmented treatments $N$ increases by one if a treatment is augmented. The outcome of interest is time to hospitalization. Hence, each patient's trajectory continues until either a failure time occurs or the study ends. A censoring variable is uniformly drawn from $[0, 25]$. When an event is censored, the trajectory ends up to the time of censoring and the censoring times are given. Here are the steps we take to generate  the dataset:

\begin{itemize}
\item Baseline variables: Variables $(\text{BP}_0, \text{weight}_0, \text{A1c}_0)$ are generated from a multivariate normal distribution with mean $(12, 140, 7.7)$ and the covariance matrix $\text{diag}(1, 1, 1)$, where BP is the systolic blood pressure. Also, $N_0 = L_0  = 0$, where $L_j$ is the discontinuation indicator at stage $j$.
\item Treatments: Given $N_j$, the sets of available treatments are  $A_{N_j=0} = \{0, \text{metformin}\}$, $A_{N_j=1} = \{0, \text{sulfonylurea}\}$, $A_{N_j=2} =
\{0,\text{glitazone}\}$, $A_{N_j=3} = \{0, \text{insulin}\}$, and $A_{N_j=4} = \{0\},$ where 0 means continue with the current treatment. The treatment is given as follows,
\begin{itemize}
\item if $\text{A1c}_j < 7$, continue with the current treatment and $N_j = N_{j-1}$.
\item if $\text{A1c}_j > 8$ and $N_{j-1} < 4$, augment the current treatment and $N_j = N_{j-1} + 1$.
\item if $7 < \text{A1c}_j < 8$ and $N_{j-1} < 4$, then a binary variable $Z_j$ is generated with probability
$$
P(Z_j = 1| \text{A1c}_{j-1}, N_{j-1}, L_{j-1}) = \frac{\exp(-0.2 \text{A1c}_{j-1} + 0.5N_{j-1} + 0.5L_{j-1})}{1+\exp(-0.2 \text{A1c}_{j-1} + 0.5N_{j-1} + 0.5L_{j-1})},
$$
where $L_j$ is the discontinuation indicator. If $Z_j = 1$, the patient continues with the current treatment, and we set $A_j = -1, N_j = N_{j-1}$. If $Z_j = 0$, the treatment is augmented, and we set $A_j = 1, N_j = N_{j - 1} + 1$.
\end{itemize}
\item Treatment discontinuation indicator: A binary variable $L_j$ is generated from a  Bernoulli distribution given the last augmented treatment. The treatment discontinuation rates are $P(L_j|A_{j-1} = \text{metformin}) = P(L_j|A_{j-1} = \text{sulfonylurea}) = P(L_j|A_{j-1} = \text{glitazone}) = 0.20$, and $P(L_j|A_{j-1} = \text{insulin}) = 0.35$.
We assume that $P(L_j = 1|A_{j-1} = 0) = 0$.
\item A1c, BP and weight at time $j$: we use the following generative model for A1c,
$$
\text{A1c}_j = \frac{\text{A1c}_{j-1} - \mu_{j - 1} + \epsilon}{\sqrt{1+\sigma_{\epsilon}^2} + \mu_j},
$$
where $\epsilon \sim N(0, 0.25)$, $\sigma_\epsilon^2 = 0.25$ and
$$
\mu_j = \left\{\begin{array}{l}
                      \mu_{j-1}(1-\tau_{A_j}) \quad \mbox{if }  \text{A1c}_{j-1} > 7, N_{j-1} < 4,  A_j \neq 0 \mbox{ and } L_j \neq 1,\\
                      \mu_{j-1} \quad \mbox{o.w.}
                      \end{array}\right. .
$$
$\tau_{A_j}$ is the treatment effect of $A_j$, where the treatment effects of metformin, sulfonylurea, glitazone  and insulin are 0.14, 0.20, 0.12, and 0.14, respectively. For the other time-varying variables at time $j$: $\text{BP}_j = (\text{BP}_{j-1} + \epsilon_1)/\sqrt{1+\sigma_{\epsilon_1}^2}$ and $\text{weight}_j = (\text{weight}_{j-1} + \epsilon_2)/\sqrt{1+\sigma_{\epsilon_2}^2}$, $\epsilon_1, \epsilon_2 \sim N(0, 0.25)$.

\item Time to hospitalization: two generative mechanisms are considered. In Scenario 1, the survival time at stage $j$, i.e., time to hospitalization, starting from the beginning of stage $j$, is generated by
$$
\log( Y_j )   =  2.5 - 0.5\times|\text{A1c}_j + 0.5N_{j-1}- 10| \{I(\text{A1c}_j>0) -  I(\text{A1c}_j + 0.5N_{j-1} > 10)\}^2 + \varepsilon,
$$
where $\varepsilon$ follows a standard normal distribution. In Scenario 2, the survival time at stage $j$ is generated by
$$
\log( Y_j )   =  2.5 - 0.5\times| \text{A1c}_j  - 7| \{I( \text{A1c}_j>0) -  I( \text{A1c}_j + 0.5N_{j-1} > 10)\}^2 + \varepsilon.
$$
\end{itemize}

The regret at each stage, i.e., the loss of reward incurred by not following the optimal treatment regime at each stage,  for Scenario 1 is $0.5\times|\text{A1c}_j + 0.5N_{j-1}- 10| \{I(\text{A1c}_j>0) -  I(\text{A1c}_j + 0.5N_{j-1} > 10)\}^2$, and the regret for Scenario 2 is $0.5\times|\text{A1c}_j  - 7| \{I(\text{A1c}_j>0) -  I(\text{A1c}_j + 0.5N_{j-1} > 10)\}^2$. Then the underlying optimal rule is the rule that yields zero regret for all stages. Hence, in both scenarios, the optimal DTR is shared across stages, and $d_j(H_j) = \sign(\text{A1c}_j + 0.5N_{j-1} > 10)$, where $H_j = (\text{A1c}_j, \text{BP}_j, \text{Weight}_j, L_j, N_{j-1})$ in our situation. In the first example, the difference between the treatment effects, also known as the contrast function, can be specified as $\text{A1c}_j + 0.5N_{j-1}- 10$. A linear model in censored Q-learning or censored shared-Q-learning could be close enough to a correctly specified model. However, this is not true in the second example, where a linear model is severely misspecified.

The proposed censored shared-Q-learning  and censored shared-O-learning   are compared with  censored Q-learning  \citep{goldberg2012q}, which does not take into account the shared data structure.  We consider sample sizes of 2000 and 5000. We generate a validation dataset with 50000 observations. The experiment is performed 500 times independently. In each replicate, we calculate the mean response for all subjects in the validation dataset, had the whole population followed the estimated rule. The averaged outcome is used to compare different methods.

The censored shared-O-learning is implemented following \eqref{emp_obj} with $\phi(t) = -\log(1+e^{-t})$, and $K$ in the soft minimum function is set to 1. We looked at other values of $K$, which gave similar results.  The propensity scores at each stage are estimated using the treatment proportion.  For the censored shared-Q-learning, we use the linear model for the Q-function, and parameters are estimated via weighted least squares as presented in \eqref{sharedQ}. For the censored Q-learning, the linear model is also used for the Q-function, where we let  $Q_j(H_j, A_j) = \beta_j^\T H_{j0} + (\psi_j^\T H_{j1})A_j$. A weighted least square is utilized at each stage to solve for $\hat\beta_j$ and $\hat\psi_j$.


We use  the Cox proportional hazards model to estimate $S_C\{\min(\sum_{k=1}^j Y_k, C) \cond H_j, A_j\}$ and construct the weight  at each stage. Note that in censored shared-O-learning method, only $S_C\{\min(\sum_{k=1}^T Y_k, C) \cond H_T, A_T\}$ needs to be fitted. For all stages, we use A1c, BP and weight at the baseline level to fit the Cox models. Let $Z_C$ denote the regressors, and $\lambda_{Ci}(t)$   denote the hazard
functions of censoring times for subject $i$. Then, $\lambda_{Ci}(t)=\lambda_{C0}(t)\exp(\beta^\T_C Z_{Ci})$, where
$\lambda_{C0}(t)$ is the baseline hazard functions for censoring time.
The estimator for $\beta_C$, say $\widehat\beta_C$, maximizes
the partial likelihood
  $$\prod_{i=1}^n \left\{\frac{\exp(\beta^\T_C Z_{Ci})}{\sum_{U_j \leq U_i}\exp(\beta^\T_C Z_{Cj})}\right\}^{1-\Delta_i}.$$
  We use the Breslow estimator  for the cumulative baseline
  hazard function $\Lambda_{C0}(t)$.
  An estimator of $S_C(t \cond H_{j i}, A_{j i})$   for subject $i$ is $\widehat S_C(t \cond H_{\bar T, i}, A_{\bar T, i}) =
  \exp\{-\widehat\Lambda_{C0}(t)\}^{\exp(\widehat\beta_C^\T Z_{Ci})}$, where $\widehat\Lambda_{C0}(t)$ is the estimator for $\Lambda_{C0}(t)$.

The means and the standard errors of the 500 values of the estimated DTRs on the validation set for  both scenarios are presented in Table \ref{tbsim1}. In Scenario 1, when the regression model is close to being correctly specified, the censored shared-Q-learning leads to the best performance. The censored Q-learning approach does not account for the shared data structure. Thus, there is a large variation in the obtained results.  In Scenario 2, both censored shared-Q-learning and censored Q-learning methods are sensitive to model misspecification. Conversely, censored shared-O-learning has a robust performance, though it has slightly worse results compared to censored shared-Q-learning method in Scenario 1.  In practice, it is unknown to us whether the regression model in censored shared-Q-learning could be correctly specified. We can use a cross-validation approach to choose the one that yields a better result (e.g. larger estimated value) between censored shared-Q-learning and censored shared-O-learning method.

\section{Data Analysis} \label{scn_data}

In this section, we apply the proposed methods to the Framingham Heart Study. The Framingham Heart Study, established in 1948, is the first longitudinal prospective large-scale cohort to study cardiovascular disease in the US.  In the original cohort 5209 men and women are monitored prospectively for epidemiological and genetic risk factors
for cardiovascular disease. There are maximum 32 examinations that occurred biannually during the 65 years of followup \citep{tsao2015cohort}.  For illustration we consider only information from the second to the sixth visits.  In our dataset, 2236 subjects are available with complete information on the risk factors at each measurement time and are free of cardiovascular disease at the time of examination. The long-term outcome of interest is time to the onset of the first major cardiovascular disease event or death, which has an event rate of 18.7\% by the end of the study. The median followup time is 25 years and the ages at baseline range from 17 to 70 with a median of 43. Traditionally, hypertension medication is recommended based on the blood pressure level. However, the outcomes might be improved if other information is also factored in. We utilize the Framingham Heart Study data to derive a decision rule that informs a patient whether a hypertension medication should be taken at each decision point, aiming to reduce the long-term risk of cardiovascular disease. Risk factors considered in our prediction at each visit include age, diastolic blood pressure, cholesterol, high-density lipoprotein, presence of diabetes, and smoking.

We first carry out a cross-validation procedure to select the method  between censored shared-Q-learning and censored shared-O-learning. At each run, we partition the whole dataset into two parts, with one part serving as training data to estimate the SDTRs using both methods, and the other part as the validation set for implementing the estimated SDTRs. When estimating SDTRs, the Kaplan--Meier method is used to fit the censoring probabilities. The constructed SDTRs from the training set are evaluated using the empirical value on the validation set adjusted for censoring.  Each part served as the validation subset once, and the cross-validated values are obtained by averaging the empirical values on both validation subsets. The procedure is repeated 100 times.   

\begin{figure}[!p]
\begin{center}
\includegraphics[width = 1.89in, height = 2in, trim =  {0  0 1.9in 0}, clip]{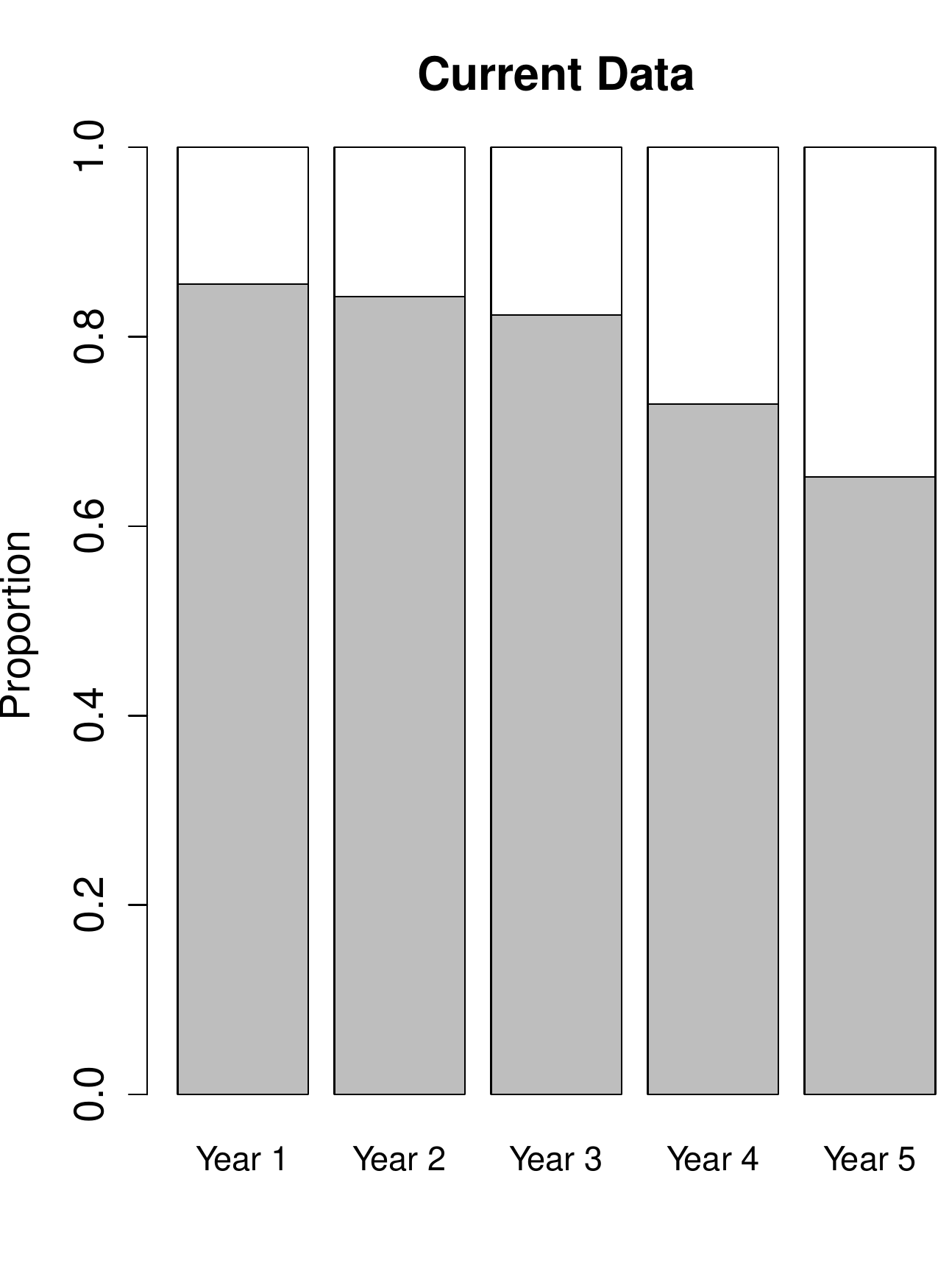}
\includegraphics[width = 1.89in, height = 2in, trim =  {0  0 1.9in 0}, clip]{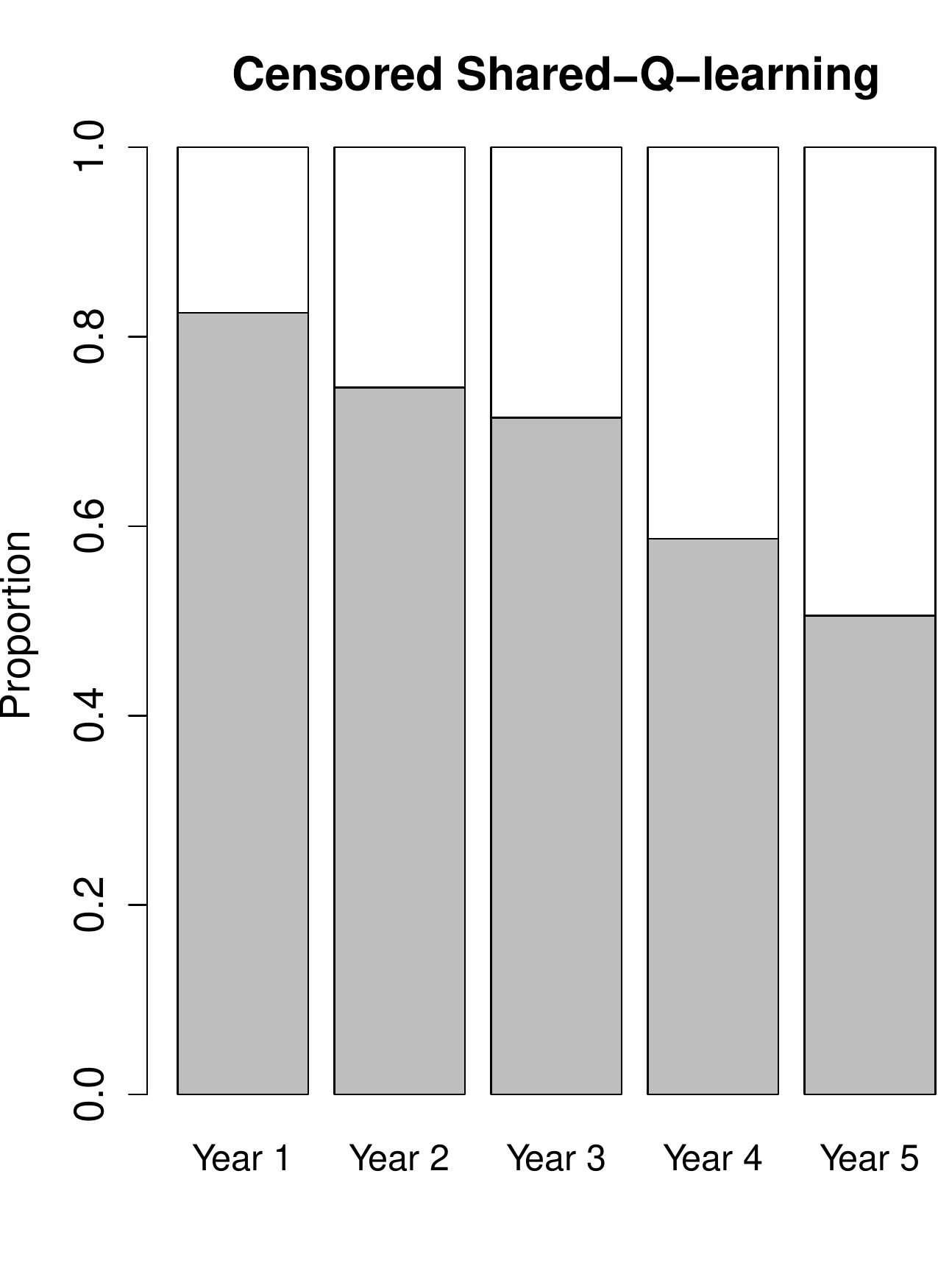}
\includegraphics[width = 2.554 in, height = 2in ]{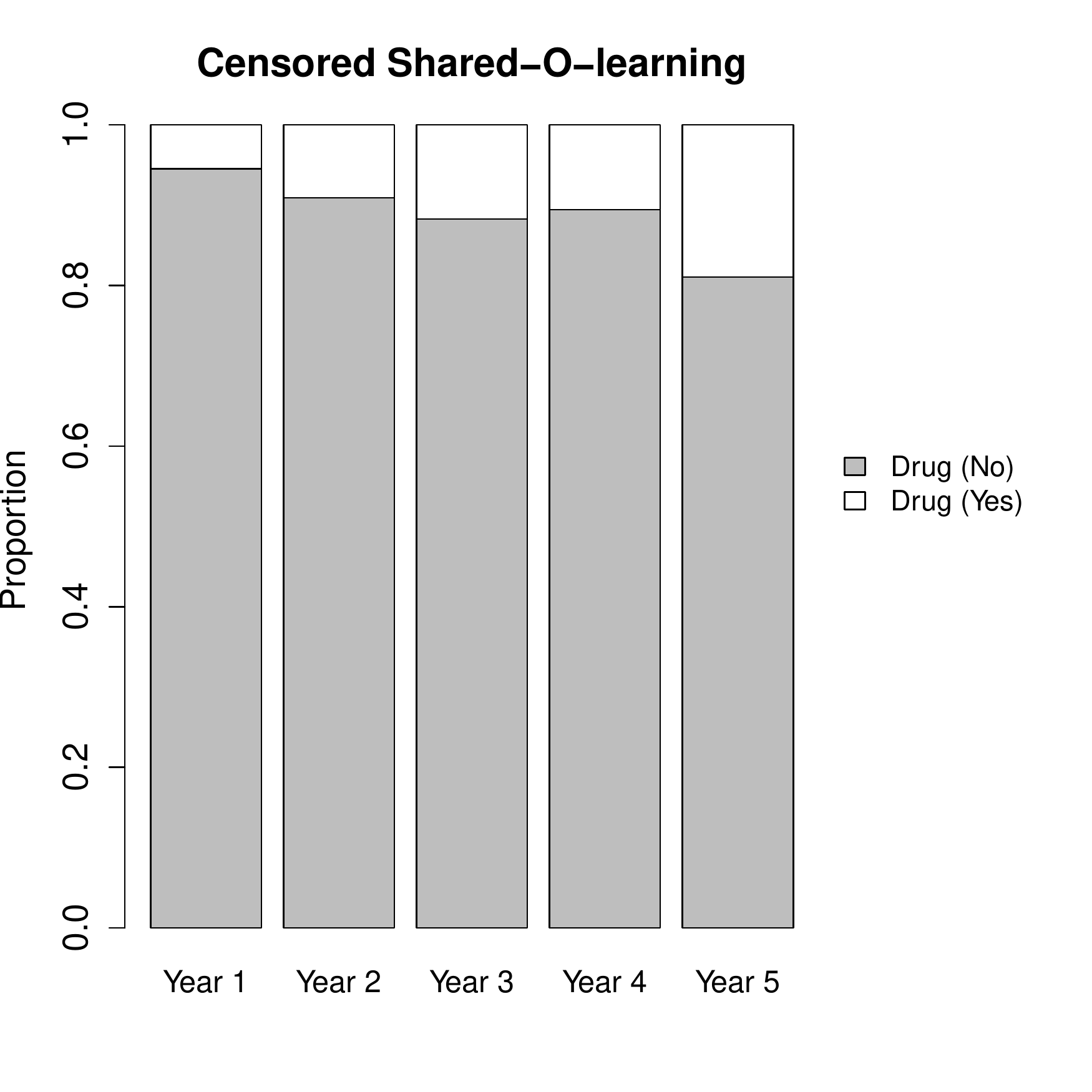}
 \end{center}
\caption{Treatment allocation of the current data, recommended rules from CSQL and CSOL; CSQL: censored shared-Q-learning; CSOL: censored shared-O-learning}
\label{txProp}
\end{figure}

\begin{figure}[!p]
\begin{center}
\includegraphics[width = 3in]{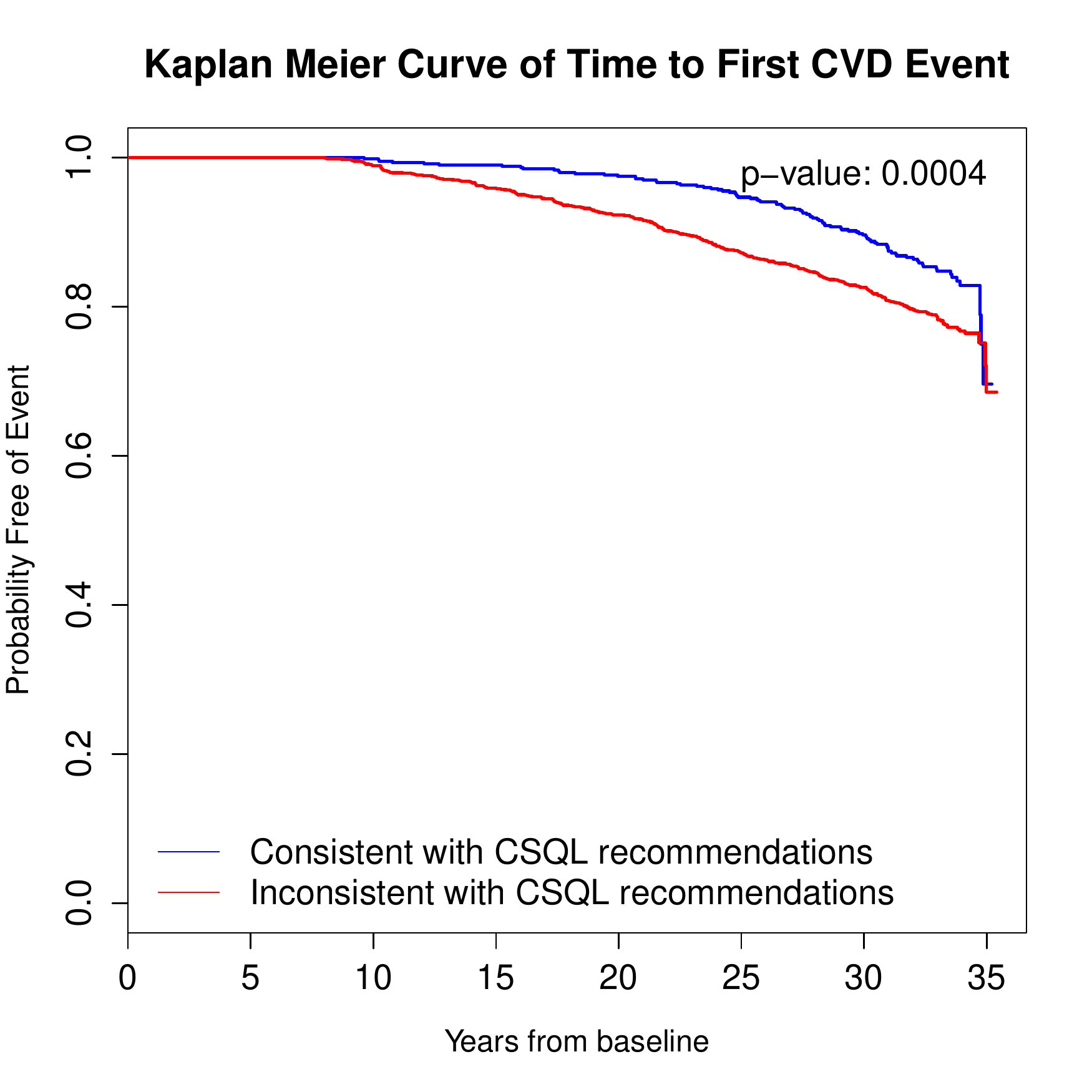}
\includegraphics[width = 3in]{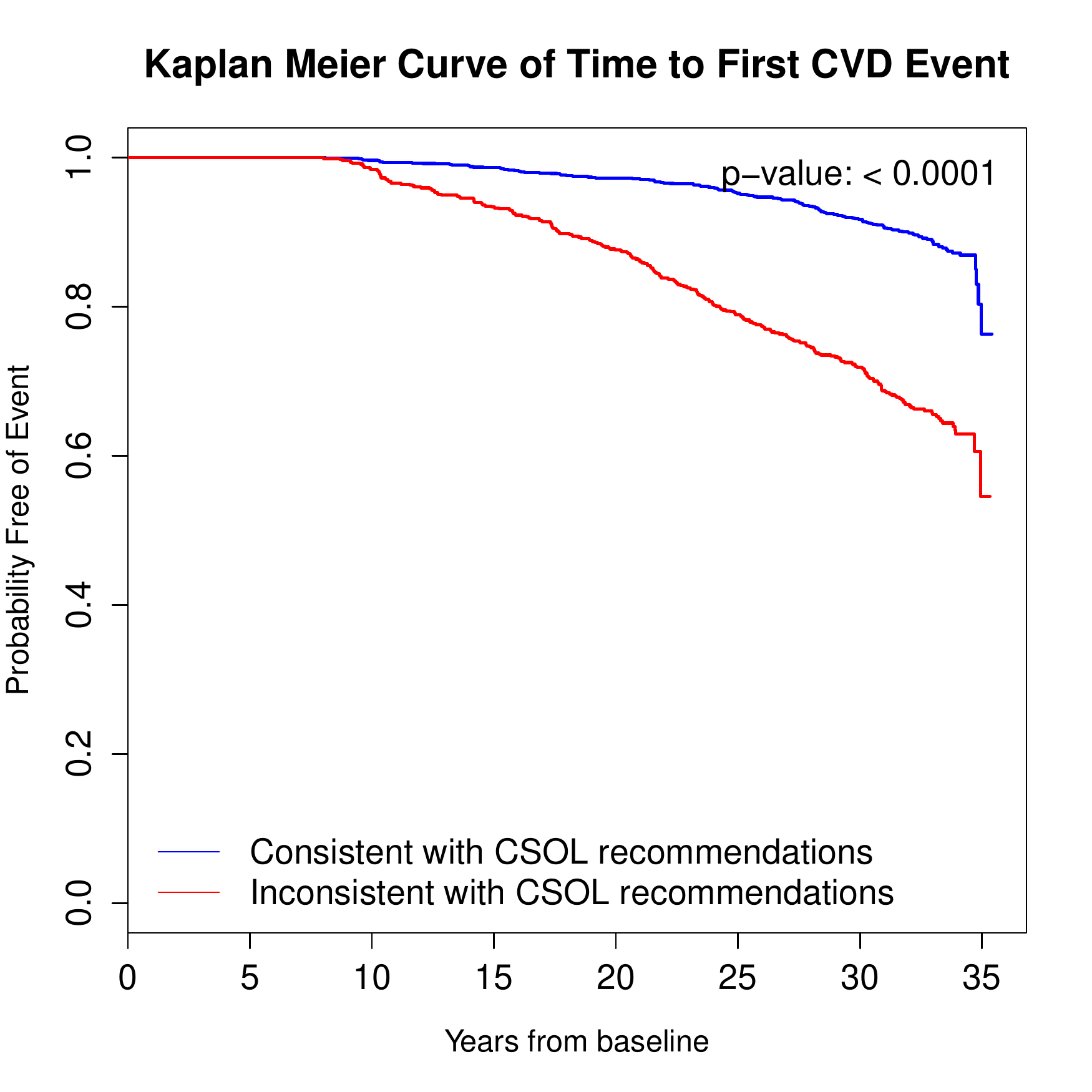}
 \end{center}
\caption{Kaplan--Meier Curves of Probability of Time to First cardiovascular disease Event (left panel: CSQL; right panel: CSOL); CSQL: censored shared-Q-learning; CSOL: censored shared-O-learning}
\label{km_FRS}
\end{figure}

Our implementation shows that both methods result  in similar cross-validated values   of mean residual survival time in years   (censored shared-Q-learning: 16.53, and censored shared-O-learning: 16.80). Hence, we carry out both methods on the whole dataset. The coefficients in the estimated SDTR  are presented in Table \ref{coef_data}, and Figure \ref{txProp} shows the treatment allocation rates from the constructed dynamic treatment regimes. The recommended rules may look different between the two methods. This could happen when there are patients who don't have great differential treatment effects. In general, patients who are currently on medication are more likely to continue taking it.  There is a slight increase in the proportion of subjects recommended for medication in the later years from censored shared-Q-learning recommendations, compared with the current data. Conversely, fewer patients are recommended with medications using censored shared-O-learning rules. In either case, the survival benefit could be significantly improved under the recommended SDTRs. 
 Figure \ref{km_FRS} shows the Kaplan-Meier curves of time to first cardiovascular disease event for patients whose treatments are consistent with censored shared-Q-learning recommendations (left panel) and censored shared-O-learning recommendations (right panel) versus those who are not, at the first and subsequent time points. It is clear that subjects whose medication coincided with the recommendation had on average better survival outcome.  

\section{Discussion} \label{scn_disc}

We proposed two new methods for constructing a SDTR with survival outcomes, which is a fixed function of time-varying covariates over time. Such a rule yields an optimal treatment strategy that can be easily implemented in practice. We provide efficient computing algorithms to obtain the solution. Our method for the decision rule is based on a linear combination of updated covariate information. It is also of interest to develop a more robust tree structured decision rule, e.g. \cite{zhu2017greedy}, without the assumption of linearity, which has the additional advantage of ease of interpretation and dissemination. The censored shared-O-learning is proposed based on an inverse probability weighted estimator of the expected outcome that would be achieved under a particular DTR. However, such an estimator is potentially less efficient because it only uses outcome information from subjects whose treatment assignments coincide with those dictated by the DTR of interest. In the future, we can develop SDTRs via censored shared-O-learning using an augmented inverse probability weighted estimator  \citep{tsiatis2006semiparametric, zhang2013robust}. Such an approach can incorporate contributions from the subjects who did not receive the specified treatment assignments across all stages by estimating their pseudo outcomes using censored Q-learning. Hence, it will improve efficiency over the censored shared-O-learning proposed here.




\bibliographystyle{jasa}
\bibliography{stabilizeddtr}

\begin{table}[!p]
\caption{Mean values (s.e.) under the estimated optimal treatment rules; Opt: optimal value; CQL: censored Q-learning; CSQL: censored shared-Q-learning; CSOL: censored shared-O-learning.}
\begin{center}
\begin{tabular}{ccccccc}
\hline\hline
Scenario & $n$ & \# of Stages & Opt & CQL & CSQL & CSOL\\
1 & 2000 & 10 & 9.71  & 7.56 (2.43) & 9.63 (0.11) & 8.25 (0.53) \\
1 & 2000 & 20 & 18.83   & 13.51 (5.83) & 18.79 (0.07) & 16.37 (1.62) \\
1 & 5000 & 10 & 9.71  & 7.75 (2.33) & 9.66 (0.05) & 8.19 (0.47) \\
1 & 5000 & 20 & 18.83   & 14.53 (5.42) & 18.81 (0.03) & 16.28 (1.70) \\
2 & 2000 & 10 & 9.68   & 5.89 (2.02) & 4.96 (0.04) & 7.86 (0.18) \\
2 & 2000 & 20 & 18.76   & 8.27 (4.26) & 5.72 (0.09) & 14.70 (0.21) \\
2 & 5000 & 10 & 9.68   & 5.38 (1.61) & 4.96 (0.03) & 7.86 (0.21) \\
2 & 5000 & 20 & 18.76 & 7.02 (3.49) & 5.73 (0.05) & 14.67 (0.71) \\
\hline\hline
\end{tabular}
\label{tbsim1}
\end{center}
\end{table}

\begin{table}[!p]
\caption{Coefficients by CSQL and CSOL for the Framingham Heart Study (CSQL: censored shared-Q-learning; CSOL: censored shared-O-learning)}
\begin{center}
\begin{tabular}{rrr}
\hline\hline
&CSQL  & CSOL\\
\hline
Intercept & 1 & -1 \\
Age &  0.020  & 0.005\\
Medication (Yes = 1) & 0.414 & 0.269 \\
Diastolic blood pressure  &-0.027  & 0.007\\
Total cholesteral  &-0.007 & -0.000\\
HDL cholesterol &0.024  & 0.001\\
Diabetes  (Yes = 1)  & -0.268 & 0.098\\
\hline\hline
\end{tabular}
\label{coef_data}
\end{center}
\end{table}

\end{document}